# Title: An Earth-sized Planet in the Habitable Zone of a Cool Star

**Authors:** Elisa V. Quintana[1,2]\*, Thomas Barclay[2,3], Sean N. Raymond[4,5], Jason F. Rowe[1,2], Emeline Bolmont[4,5], Douglas A. Caldwell[1,2], Steve B. Howell[2], Stephen R. Kane[6], Daniel Huber[1,2], Justin R. Crepp[7], Jack J. Lissauer[2], David R. Ciardi[8], Jeffrey L. Coughlin[1,2], Mark E. Everett[9], Christopher E. Henze[2], Elliott Horch[10], Howard Isaacson[11], Eric B. Ford[12,13], Fred C. Adams[14,15], Martin Still[3], Roger C. Hunter[2], Billy Quarles[2], Franck Selsis[4,5]

**Affiliations:**

[1]SETI Institute, 189 Bernardo Ave, Suite 100, Mountain View, CA 94043, USA.

[2]NASA Ames Research Center, Moffett Field, CA 94035, USA.

[3]Bay Area Environmental Research Institute, 596 1st St West Sonoma, CA 95476, USA.

[4]Univ. Bordeaux, Laboratoire d'Astrophysique de Bordeaux, UMR 5804, F-33270, Floirac, France.

[5]CNRS, Laboratoire d'Astrophysique de Bordeaux, UMR 5804, F-33270, Floirac, France.

[6]San Francisco State University, 1600 Holloway Avenue, San Francisco, CA 94132, USA.
[7]University of Notre Dame, 225 Nieuwland Science Hall, Notre Dame, IN 46556, USA.
[8]NASA Exoplanet Science Institute, California Institute of Technology, 770 South Wilson Avenue Pasadena, CA 91125, USA.
[9]National Optical Astronomy Observatory, 950 N. Cherry Ave, Tucson, AZ 85719
[10]Southern Connecticut State University, New Haven, CT 06515
[11]University of California, Berkeley, CA, 94720, USA.
[12]Center for Exoplanets and Habitable Worlds, 525 Davey Laboratory, The Pennsylvania State University, University Park, PA, 16802, USA
[13]Department of Astronomy and Astrophysics, The Pennsylvania State University, 525 Davey Laboratory, University Park, PA 16802, USA
[14]Michigan Center for Theoretical Physics, Physics Department, University of Michigan, Ann Arbor, MI 48109, USA
[15]Astronomy Department, University of Michigan, Ann Arbor, MI 48109, USA

\*Correspondence to: elisa.quintana@nasa.gov

**Abstract**:

The quest for Earth-like planets is a major focus of current exoplanet research. Although planets that are Earth-sized and smaller have been detected, these planets reside in orbits that are too close to their host star to allow liquid water on their surfaces. We present the detection of Kepler-186f, a $1.11\pm0.14$ Earth radius planet that is the outermost of five planets, all roughly Earth-sized, that transit a $0.47\pm0.05$ solar radius star. The intensity and spectrum of the star's radiation places Kepler-186f in the stellar habitable zone, implying that if Kepler-186f has an Earth-like atmosphere and water at its surface, then some of this water is likely to be in liquid form.

**Main Text:**

In recent years we have seen great progress in the search for planets that, like our own, are capable of harboring life. Dozens of known planets orbit within the habitable zone (HZ), the region around a star within which a planet can sustain liquid water on its surface (*1-4*). Most of these HZ planets are gas giants, but a few, such as Kepler-62f (*5*), are potentially rocky despite having a larger radius than Earth. Hitherto, the detection of an Earth-sized planet in the HZ of a main-sequence star has remained elusive.

Low-mass stars are good targets in the search for habitable worlds. They are less luminous than the Sun so their HZs are located closer in (*6*). The shorter orbital period and larger planet-to-star size ratio of a planet in the HZ of a cool star relative to planets orbiting in the HZ of solar-type stars allow for easier transit detections. M-dwarfs, stars with 0.1-0.5 times the mass of the Sun ($M_\odot$), are very abundant, comprising about three quarters of all main sequence stars in our galaxy (*7*). They also evolve very slowly in luminosity, thus their habitable zones remain nearly constant for billions of years.

Kepler-186 (also known as KIC8120608 and KOI-571) is a main-sequence M1-type dwarf star with a temperature of 3788+/-54 K and an iron abundance half that of the Sun (*8* and SOM Section 2). The star was observed by the Kepler spacecraft at near-continuous 29.4-min intervals. Four planets designated Kepler-186b-e, all smaller than 1.5 $R_\oplus$ with orbital periods between 3.9 and 22.4 days, were confirmed with the first two years of data (*9, 10*). The fifth planet candidate, Kepler-186f which we discuss herein, was detected with an additional year of data (*14*).

We compared the observed data to a five planet model with limb-darkened transits (*9, 11*) allowing for eccentric orbits to estimate the physical properties of Kepler-186f. We used an affine invariant Markov-chain Monte Carlo (MCMC) algorithm (*12, 13*) to efficiently sample the model parameter posterior distribution. Kepler-186f has an orbital period of 129.9 days and a planet-to-star radius ratio of 0.021. The additional constraint on stellar density from the transit model allowed us to refine the stellar radius that was previously derived by modeling spectroscopic data. Interior models of cool main-sequence stars such as Kepler-186 show systematic differences to empirically measured stellar properties (*14-16*, SOM Section 2). To account for discrepancies between the empirically measured radii and those derived from model isochrones at the measured temperature for Kepler-186, we have added a 10% uncertainty in quadrature to our stellar radius and mass estimate, yielding a final estimate of $R_\star$ = 0.472+/-0.052 and a planet radius of 1.11±0.14 $R_\oplus$ (Fig. 1, Table S2).

The Kepler-186 planets do not induce a detectable reflex motion on the host star or dynamically perturb each other so as to induce substantial non-Keplerian transit ephemerides, both of which can be used to help confirm the planetary nature of Kepler's planet candidates (*17, 18*). Instead, we used a statistical approach to measure the confidence in the planetary interpretation of each candidate planet (*19, 20*). We obtained follow-up high-contrast imaging observations using the Keck-II and Gemini-North telescopes (SOM Section 5) to restrict the parameter space of stellar magnitude/separation where a false positive inducing star could reside and mimic a planetary transit. No nearby sources were observed in either the Keck-II or Gemini data; the 5-σ detection limit set the brightness of a false-positive star to be *Kp*=21.9 at 0.5″ from Kepler-186 and 19.5 at 0.2″ where *Kp* is the apparent magnitude of a star in the Kepler bandpass.

The probability of finding a background eclipsing binary or planet hosting star that could mimic a transit in the parameter space not excluded by observations is very low: 0.5% chance relative to the probability that we observe a planet orbiting the target. However, this does not account for the possibility that the planets orbit a fainter bound stellar companion to Kepler-186. Although we have no evidence of any binary companion to the target star, faint unresolved stellar companions to planet host stars do occur (*21*). We constrained the density of the host star from the transit model by assuming that all five planets orbit the same star. The 3-σ upper bound of the marginalized probability density function of stellar density from our MCMC simulation is 11.2 g cm$^{-3}$. If Kepler-186 and a hypothetical companion co-evolved, the lower limit on the stellar mass and brightness of a companion would be 0.39 $M_\odot$ and *K*p=15.1, respectively.

Given the distance to Kepler-186 of 151±18 pc, a companion would have to be within a projected distance of 4.2 AU from the target to avoid detection via our follow-up observations. However, a star closer than 1.4 AU from the primary would cause planets around the fainter star to become unstable (*22*). The probability of finding an interloping star with the specific parameters needed to masquerade as a transiting planet is very small relative to the a priori probability that the planets orbit Kepler-186 (<0.02%). Therefore we are confident that all five planets orbit Kepler-186.

While photometry alone does not yield planet masses, we used planetary thermal evolution models to constrain the composition of the Kepler-186 planets. These theories predict that the composition of planets with radii less than about 1.5 $R_\oplus$ are unlikely to be dominated by H/He gas envelopes (*23*). Although a thin H/He envelope around Kepler-186f cannot be entirely ruled out, the planet was likely vulnerable to photo-evaporation early in the star's life when extreme ultra-violet (XUV) flux from the star was significantly higher. Hence any H/He envelope that was accreted would likely have been stripped via hydrodynamic mass loss (*23*). Although Kepler-186f likely does not have a thick H2-rich atmosphere, a degeneracy remains between the relative amounts of iron, silicate rock, and water since the planet could hold on to all of these cosmically-abundant constituents. Mass estimates for Kepler-186f can therefore range from 0.32 $M_\oplus$ if composed of pure water/ice, 3.77 $M_\oplus$ if the planet is pure iron, and an Earth-like composition (about 1/3 iron and 2/3 silicate rock) would give an intermediate mass of 1.44 $M_\oplus$ (Table S3).

For Kepler-186, the conservative estimate of the habitable zone (i.e., likely narrower than the actual annulus of habitable distances) extends from 0.22-0.40 AU (*4*). The four inner planets are too hot to ever enter the habitable zone. Kepler-186f receives $32^{+6}_{-4}$% of the intensity of stellar radiation (i.e., insolation) as that received by Earth from the Sun. Despite receiving less energy than the Earth, Kepler-186f is within the habitable zone throughout its orbit (Fig. 2). It is difficult for an Earth-size planet in the habitable zone of an M star to accrete and retain $H_2O$ (*24*, *25*), but being in the outer portion of its star's habitable zone reduces these difficulties.

The high coplanarity of the planets' orbits (given by the fact that they all transit the star) suggest that they formed from a protoplanetary disk. The leading theories for the growth of planets include in-situ accretion of local material in a disk (*26*, *27*), collisional growth of inward-migrating planetary embryos (*28*, *29*), or some combination thereof. We performed a suite of *N*-body simulations of late-stage in situ accretion from a disk of planetary embryos around a star like Kepler-186 (SOM Section 9). We found that a massive initial disk (>10 $M_\oplus$) of solid material with a very steep surface density profile is needed to form planets similar to the Kepler-186 system. Accretion disks with this much mass so close to their star (< 0.4 AU) or with such steep surface density profiles, however, are not commonly observed (*30*), suggesting that the Kepler-186 planets either formed from material that underwent an early phase of inward migration while gas was still present in the disk (*31*) or were somehow perturbed inwards after they formed. Regardless, all simulations produced at least one stable planet in between the orbits of planets e and f, in the range 0.15-0.35 AU (Fig. S5). The presence of a sixth planet orbiting between e and f is not excluded by the observations; if such a planet were to have a modest inclination of a few degrees with respect to the common plane of the other planets we would not observe a transit.

Planets that orbit close to their star are subjected to tidal interactions that can drive the planets to an equilibrium rotational state, typically either a spin-orbit resonance or a "pseudo-synchronous" state whereby the planet co-rotates with the star at its closest approach (*32*, *33*). The proximity of the inner four planets to Kepler-186 suggests that they are likely tidally locked. Kepler-186f, however, is at a large enough distance from the star such that uncertainties in the tidal dissipation function precludes any determination of its rotation rate (*34*). Regardless, tidal locking (or pseudo-synchronous rotation) does not preclude a planet from being habitable. The 5.6 Earth-mass planet GJ 581d (*35*) likely rotates pseudo-synchronously with its star and in addition receives a similar insolation (27%) as Kepler-186f.

Detailed climate models have shown GJ 581d to be capable of having liquid water on its surface (*36, 37*). Taken together, these considerations suggest that the newly discovered planet Kepler-186f is likely to have the properties required to maintain reservoirs of liquid water.

**References and Notes:**
1. S. R. Kane, D. M. Gelino, The habitable zone gallery. *Publ. Astron. Soc. Pac.*, **124**, 323-328. (2012).
2. J. F. Kasting, D. P. Whitmire, R. T. Reynolds, Habitable zones around main sequence stars. *Icarus*, **101**, 108-128. (1993).
3. F. Selsis, J. F. Kasting, B. Levrard, J. Paillet, I. Ribas, et al., Habitable planets around the star Gliese 581? *Astron. Astrophys.*, **476**, 1373-1387. (2007).
4. R. K. Kopparapu, R. Ramirez, J. F. Kasting, V. Eymet, T. D. Mahadevan, et al., Habitable zones around main-sequence stars: new estimates. *Astrophys. J.*, **765**, 131 (2013).
5. W. J. Borucki, E. Agol, F. Fressin, L. Kaltenegger, J. Rowe, et al., Kepler-62: A five-planet system with planets of 1.4 and 1.6 Earth radii in the habitable zone. *Science*, **340**, 587-590. (2013).
6. J. C. Tarter, P. R. Backus, R. L. Mancinelli, J. M. Aurnou, D. E. Backman, et al., A reappraisal of the habitability of planets around M dwarf stars. *Astrobiology*, **7**, 30-65. (2007).
7. P. Kroupa, C. A. Tout, G. Gilmore, The distribution of low-mass stars in the Galactic disc. *Mon. Not. R. Astron. Soc.*, **262**, 545-587 (1993).
8. P. S. Muirhead, K. Hamren, E. Schlawin, B. Rojas-Ayala, K. R. Covey, et al., Characterizing the cool Kepler Objects of Interests. New effective temperatures, metallicities, masses, and radii of low-mass Kepler planet-candidate host stars. *Astrophys. J.*, **750**, L37 (2012).
9. J. F. Rowe, S. T. Bryson, G. W. Marcy, J. J. Lissauer, D. Jontof-Hutter, et al., Validation of Kepler's multiple planet candidates. III: Light curve analysis & announcement of hundreds of new multi-planet systems. *Astrophy. J.*, **784**, 45 (2014).
10. J. J. Lissauer, G. W. Marcy, S. T. Bryson, J. F. Rowe, D. Jontof-Hutter, et al., Validation of Kepler's multiple planet candidates. II: Refined statistical framework and systems of special interest. *Astrophy. J.*, **784**, 44 (2014).
11. K. Mandel, E. Agol, Analytic light curves for planetary transit searches. *Astrophys. J.*, **580**, L171-L175. (2002).
12. J. Goodman, J. Weare, Ensemble samplers with affine invariance, *Comm. App. Math. Comp. Sci.*, **5**, 65. (2010).
13. D. Foreman-Mackey, D. Hogg, D. Lang, J. Goodman, emcee: The MCMC Hammer. *Publ. Astron. Soc. Pac.*, **125**, 306-312. (2013).
14. M. Lopez-Morales, On the correlation between the magnetic activity levels, metallicities, and radii of low-mass stars. *Astrophys. J.*, **660**, 732-739. (2007).

15. A. J. Bayless, J. A. Orosz, 2MASS J05162881+2607387: A new low-mass double-lined eclipsing binary. *Astrophys. J.*, **651**, 1155-1165. (2006).
16. J. Irwin, D. Charbonneau, Z. K. Berta, S. N. Quinn, D. W. Latham, et al., GJ 3236: A new bright, very low mass eclipsing binary system discovered by the MEARTH observatory. *Astrophys. J.*, **701**, 1436-1449. (2009).
17. Mayor, M., Queloz, D., A Jupiter-mass companion to a solar-type star. *Nature*, **378**, 355-359. (1995).
18. M. J. Holman, D. C. Fabrycky, D. Ragozzine, E. B. Ford, J. H. Steffen, et al., Kepler-9: A system of multiple planets transiting a sun-like Star, confirmed by timing variations. *Science*, **330**, 51. (2010).
19. G. Torres, M. Konacki, D. Sasselov, S. Jha, Testing blend scenarios for extrasolar transiting planet candidates. I. OGLE-TR-33: A false positive, *Astrophys. J.,* **614**, 979-989 (2004)
20. T. Barclay, C. J. Burke, S. B. Howell, J. F. Rowe, D. Huber, et al., A super-Earth-sized planet orbiting in or near the habitable zone around a Sun-like star. *Astrophys. J.*, **768**, 101. (2013).
21. J. Wang, J. Xie, T. Barclay, D. Fischer, Influence of stellar multiplicity on planet formation. I. Evidence of Suppressed Planet Formation due to Stellar Companions within 20 AU and Validation of Four Planets from the Kepler Multiple Planet Candidates. *Astrophys. J.*, **783**, 4. (2014).
22. E.-M. David, E. V. Quintana, M. Fatuzzo, F. C. Adams, Dynamical stability of Earth-like planetary orbits in binary systems. *Publ. Astron. Soc. Pac.*, **115**, 825-836. (2003).
23. E. D. Lopez, J. J. Fortney, N. Miller, How thermal evolution and mass-loss sculpt populations of super-Earths and sub-Neptunes: Application to the Kepler-11 system and beyond. *Astrophys. J.*, **761**, 59. (2012).
24. J. J. Lissauer, Planets formed in habitable zones of M dwarf stars probably are deficient in volatiles. *Astrophys. J.*, **660**, L149-L152. (2007).
25. S. N. Raymond, J. Scalo, V. S. Meadows, A decreased probability of habitable planet formation around low-mass stars. *Astrophys. J.*, **669**, 606-614. (2007).
26. S. N. Raymond, R. Barnes, A. M. Mandell, Observable consequences of planet formation models in systems with close-in terrestrial planets. *Mon. Not. R. Astron. Soc.*, **384**, 663-674. (2008).
27. E. Chiang, G. Laughlin, The minimum-mass extrasolar nebula: in situ formation of close-in super-Earths. *Mon. Not. R. Astron. Soc.*, **431**, 3444-3455. (2013).
28. C. Terquem, J. C. B. Papaloizou, Migration and the formation of systems of hot super-Earths and Neptunes. *Astrophys. J.,*, **654**, 1110-1120. (2007).
29. C. Cossou, S. N. Raymond, A. Pierens, Making systems of Super Earths by inward migration of planetary embryos. *IAU Symposium*, **299**, 360-364. (2014)
30. S. M. Andrews, J. P. Williams, A submillimeter view of circumstellar dust disks in ρ-Ophiuchi. *Astrophys. J.*, **671**, 1800-1812. (2007).
31. S. N. Raymond, C. Cossou, No universal minimum-mass extrasolar nebula: Evidence against in-situ accretion of systems of hot super-Earths. *Mon. Not. R. Astron. Soc.* (2014).
32. P. Hut, Tidal evolution in close binary systems. *Astron. Astrophys.*, **99**, 126-140. (1981).
33. S. Ferraz-Mello, A. Rodriguez, H. Hussmann, Tidal friction in close-in satellites and exoplanets: The

Darwin theory re-visited. *Celest. Mech. Dyn. Astron.*, **101**, 171-201. (2008).
34. R. Heller, J. Leconte, R. Barnes, Tidal obliquity evolution of potentially habitable planets. *Astron. Astrophys.*, **528**, 16. (2011).
35. S. Udry, X. Bonfils, X. Delfosse, T. Forveille, M. Mayor, et al., The HARPS search for southern extra-solar planets. XI. Super-Earths (5 and 8 M) in a 3-planet system. *Astron. Astrophys.*, **469**, L43-L47. (2007).
36. P. von Paris, S. Gebauer, M. Godolt, J. L. Grenfell, P. Hedelt, et al., The extrasolar planet Gliese 581d: a potentially habitable planet? *Astron. Astrophys.*, **522**, A23. (2010).
37. R. D. Wordsworth, F. Forget, F. Selsis, E. Millour, B. Charnay, et al., Gliese 581d is the first discovered terrestrial-mass exoplanet in the habitable zone. *Astrophys. J.* **733**, L48 (2011).
38. B. Rojas-Ayala, K. R. Covey, P. S. Muirhead, J. P. Lloyd, Metallicity and temperature indicators in M dwarf K-band spectra: Testing new and updated calibrations with observations of 133 solar neighborhood M dwarfs. *Astrophys. J.*, **748**, 93. (2012).
39. A. Dotter, B. Chaboyer, D. Jevremovic, V. Kostov, E. Baron, et al., The Dartmouth Stellar Evolution Database. *Astrophys. J.*, **178**, 89-101. (2008).
40. G. Chabrier, J. Gallardo, I. Baraffe, Evolution of low-mass star and brown dwarf eclipsing binaries. *Astron. Astrophys.*, **472**, L17-L20. (2007).
41. J. C. Morales, J. Gallardo, I. Ribas, C. Jordi, I. Baraffe, The effect of magnetic activity on low-mass stars in eclipsing binaries. *Astrophys. J.*, **718**, 502-512. (2010).
42. J. L. Coughlin, M. Lopez-Morales, T. E. Harrison, N. Ule, D. I. Hoffman, Low-mass eclipsing binaries in the initial Kepler data release. *Astron. J.*, **141**, 78. (2011).
43. A. L. Kraus, R. A. Tucker, M. I. Thompson, E. R. Craine, L. A. Hillenbrand, The mass-radius(-rotation?) relation for low-mass stars. *Astrophys. J.*, **728**, 48. (2011).
44. T. S. Boyajian, K. von Braun, G.van Belle, C. Farrington, G. Schaefer, et al., Stellar diameters and temperatures. III. Main-sequence A, F, G, and K Stars: additional high-precision measurements and empirical relations. *Astrophys. J.*, **771**, 40. (2013).
45. A. W. Mann, E. Gaidos, M. Ansdell, Spectro-thermometry of M dwarfs and their candidate planets: too hot, too cool, or just right? *Astrophys. J.*, **779**, 188. (2013).
46. T. S. Boyajian, K. von Braun, G. van Belle, H. A. McAlister, T. A. ten Brummelaar, et al., Stellar diameters and temperatures. II. Main-sequence K- and M-stars. *Astrophys. J.*, **757**, 112. (2012).
47. M. C. Stumpe, J. C. Smith, J. E. Van Cleve, J. D. Twicken, T. S. Barclay, et al., Kepler presearch data conditioning I - Architecture and algorithms for error correction in Kepler light curves. *Publ. Astron. Soc. Pac.*, **124**, 985-999. (2012).
48. J. C. Smith, M. C. Stumpe, J. E. Van Cleve, J. M. Jenkins, T. S. Barclay, et al., Kepler presearch data conditioning II - A bayesian approach to systematic error correction. *Publ. Astron. Soc. Pac.*, **124**, 1000-1014. (2012).
49. T. Barclay, M. Still, J. M. Jenkins, S. B. Howell, R. M. Roettenbacher, Serendipitous Kepler observations of a background dwarf nova of SU UMa type. *Mon. Not. R. Astron. Soc.*, **422**, 1219-1230 (2012).

50. J. Eastman, B. S. Gaudi, E. Agol, EXOFAST: A fast exoplanetary fitting suite in IDL. *Publ. Astron. Soc. Pac.*, **125**, 83-112. (2013).
51. A. Claret, S. Bloemen, Gravity and limb-darkening coefficients for the Kepler, CoRoT, Spitzer, uvby, UBVRIJHK, and Sloan photometric systems. *Astron. Astrophys.*, **529**, A75. (2011).
52. C. J. Burke, P. R. McCullough, J. A. Valenti, D. Long, C. M. Johns-Krull, et al., XO-5b: A transiting Jupiter-sized planet with a 4 day period. *Astrophys. J.*, **686**, 1331-1340. (2008).
53. W. J. Borucki, D. G. Koch, G. Basri, N. Batalha, T. M. Brown, et al., Characteristics of planetary candidates observed by Kepler. II. Analysis of the first four months of data. *Astrophys. J.*, **736**, 19. (2011).
54. N. M. Batalha, J. F. Rowe, S. T. Bryson, T. Barclay, C. J. Burke, et al., Planetary candidates observed by Kepler. III. Analysis of the first 16 months of data. *Astrophys. J. Suppl. Ser.*, **204**, 24. (2013).
55. S. T. Bryson, J. M. Jenkins, R. L. Gilliland, J. D. Twicken, B. Clarke, et al., Identification of background false positives from Kepler data. *Publ. Astron. Soc. Pac.*, **125**, 889-923. (2013).
56. S. B. Howell, J. F. Rowe, S. T. Bryson, S. N. Quinn, G. W. Marcy, et al., Kepler-21b: A 1.6 $R_\oplus$ planet transiting the bright oscillating F subgiant star HD 179070. *Astrophys. J.*, **746**, 123. (2012).
57. L. Girardi, M. Barbieri, M. A. T. Groenewegen, P. Marigo, A. Bressan, et al., in *Red Giants as Probes of the Structure and Evolution of the Milky Way*, A. Miglio, J. Montalban, A. Noels, Eds. (Springer-Verlag Berlin Heidelberg, 2012), pp. 165.
58. F. Fressin, G. Torres, D. Charbonneau, S. T. Bryson, J. Christiansen, et al., The false positive rate of Kepler and the occurrence of planets. *Astrophys. J.*, **766**, 81. (2013).
59. T. Barclay, J. F. Rowe, J. J. Lissauer, D. Huber, F. Fressin, et al., A sub-Mercury-sized exoplanet. *Nature*, **494**, 452-454. (2013).
60. D. Raghavan, H. A. McAlister, T. J. Henry, D. W. Latham, G. W. Marcy, et al., A survey of stellar families: Multiplicity of solar-type stars. *Astrophys. J. Suppl. Ser.*, **190**, 1-42. (2010).
61. J. J. Lissauer, G. W. Marcy, J. F. Rowe, S. T. Bryson, E. Adams, et al., Almost all of Kepler's multiple-planet candidates are planets. *Astrophys. J.*, **750**, 112. (2012).
62. C. Marchal, G. Bozis, Hill stability and distance curves for the general three-body problem. *Celestial Mechanics*, **26**, 311-333. (1982).
63. B. Gladman, Dynamics of systems of two close planets. *Icarus*, **106**, 247. (1993).
64. J. E. Chambers, G. W. Wetherill, A. P. Boss, The stability of multi-planet systems. *Icarus*, **119**, 261-268. (1996)
65. A. W. Smith, J. J. Lissauer, Orbital stability of systems of closely-spaced planets. *Icarus*, **201**, 381-394. (2009).
66. A. McQuillan, T. Mazeh, S. Aigrain, Stellar rotation periods of the Kepler Objects of Interest: A dearth of close-in planets around fast rotators. *Astrophys. J.*, **775**, L11. (2013).
67. L. M. Walkowicz, G. S. Basri, Rotation periods, variability properties and ages for Kepler exoplanet candidate host stars. *Mon. Not. R. Astron. Soc.*, **436**, 1883-1895. (2013).
68. N. Pizzolato, A. Maggio, G. Micela, S. Sciortino, in *Stellar Coronae in the Chandra and*

*XMM-NEWTON Era*, F. Favata, J. J. Drake, Eds. vol. 277 of Astronomical Society of the Pacific Conference Series. 557 (2002).

69. N. M. Silvestri, S. L. Hawley, T. D. Oswalt, The chromospheric activity and ages of M dwarf stars in wide binary systems. *Astron. J.*, **129**, 2428-2450. (2005).
70. J. J. Fortney, M. S. Marley, J. W. Barnes, Planetary radii across five orders of magnitude in mass and stellar insolation: Application to transits, *Astrophys. J.*, **659**, 1661-1672. (2007).
71. A. Morbidelli, J. I. Lunine, D. P. O'Brien, S. N. Raymond, K. J. Walsh, Building Terrestrial Planets, *Annu. Rev. Earth Planet. Sci.*, **40**, 251-275. (2012).
72. J. E. Chambers, A hybrid symplectic integrator that permits close encounters between massive bodies. *Mon. Not. R. Astron. Soc.*, **304**, 793-799. (1999).
73. D. Valencia, D. D. Sasselov, R. J. O'Connell, Detailed models of super-Earths: How well can we infer bulk properties? *Astrophys. J.*, **665**, 1413-1420. (2007).
74. P. Tenenbaum, J. M. Jenkins, S. Seader, C. J. Burke, J. L. Christiansen, et al., Detection of potential transit signals in the first 12 quarters of Kepler Mission data. *Astrophys. J.*, **206**, 5. (2013).

**Acknowledgments:**

The authors working at NASA Ames would like to thank the SETI Institute for hosting them during the US government shutdown. E.V.Q. and J.F.R. acknowledge support from ROSES Kepler Participating Scientist Program Grant NNX12AD21G. S.N.R.'s contribution was performed as part of the NASA Astrobiology Institute's Virtual Planetary Laboratory Lead Team, supported by the NASA under Cooperative Agreement No. NNA13AA93A. D.H. acknowledges support by an appointment to the NASA Postdoctoral Program at Ames Research Center, administered by Oak Ridge Associated Universities through a contract with NASA, and the Kepler Participating Scientist Program. The Center for Exoplanets and Habitable Worlds is supported by the Pennsylvania State University, the Eberly College of Science, and the Pennsylvania Space Grant Consortium. F. S. acknowledges support from the European Research Council (Starting Grant 209622: E3ARTHs). This paper includes data collected by the Kepler mission. Funding for the Kepler mission is provided by the NASA Science Mission directorate. This research has also made use of NASA's Astrophysics Data System. Some of the data presented in this paper were obtained from the Mikulski Archive for Space Telescopes (MAST).

STScI is operated by the Association of Universities for Research in Astronomy, Inc., under NASA contract NAS5-26555. Support for MAST for non-HST data is provided by the NASA Office of Space Science via grant NNX13AC07G and by other grants and contracts. This research has made use of the NASA Exoplanet Archive, which is operated by the California Institute of Technology, under contract with the National Aeronautics and Space Administration under the Exoplanet Exploration Program. Gemini Observatory is operated by the Association of Universities for Research in Astronomy, Inc., under a cooperative agreement with the NSF on behalf of the Gemini partnership: the National Science Foundation (United States), the National Research Council (Canada), CONICYT (Chile), the Australian Research Council (Australia), Ministério da Ciência, Tecnologia e Inovação (Brazil) and Ministerio de Ciencia, Tecnología e Innovación Productiva (Argentina).

**Fig. 1**. The five transiting planet signals observed by Kepler, folded on the orbital periods of the respective planets. The plots are ordered by ascending planet orbital periods. The black points show the observed data and the blue points are the observed data binned in time with one point per phase-folded hour. The most probable transit model is shown in red. The incomplete phase coverage for Kepler-186d is a result of the orbital period of the planet having a value close to an integer multiple of the sampling.

**Fig. 2**. A schematic diagram of the Kepler-186 system. The upper section of the plot shows a top-down view of the system during a transit of planet f. The relative planet sizes are correct but are not on the same scale as the orbits (shown as black curves). The lower section shows a side-on view comparing Kepler-186 with the solar system (with Earth and Mars in the habitable zone) and the Gliese 581 planets. The stars are located at the left edge of the plot. The dark grey regions represent conservative estimates of the habitable zone while the lighter grey regions are more optimistic extensions of the habitable region around each star (*3*, *4*). Kepler-186f receives $0.32^{+0.06}_{-0.04}$ of the incident flux that the Earth receives from the Sun. This puts Kepler-186f comfortably within the conservative HZ, which ranges from 0.25 to 0.88 of Earth's incident flux for this star. Kepler-186f receives a similar incident flux to Gliese 581d (*35*) which has been shown to be capable of hosting liquid water (*36*, *37*).

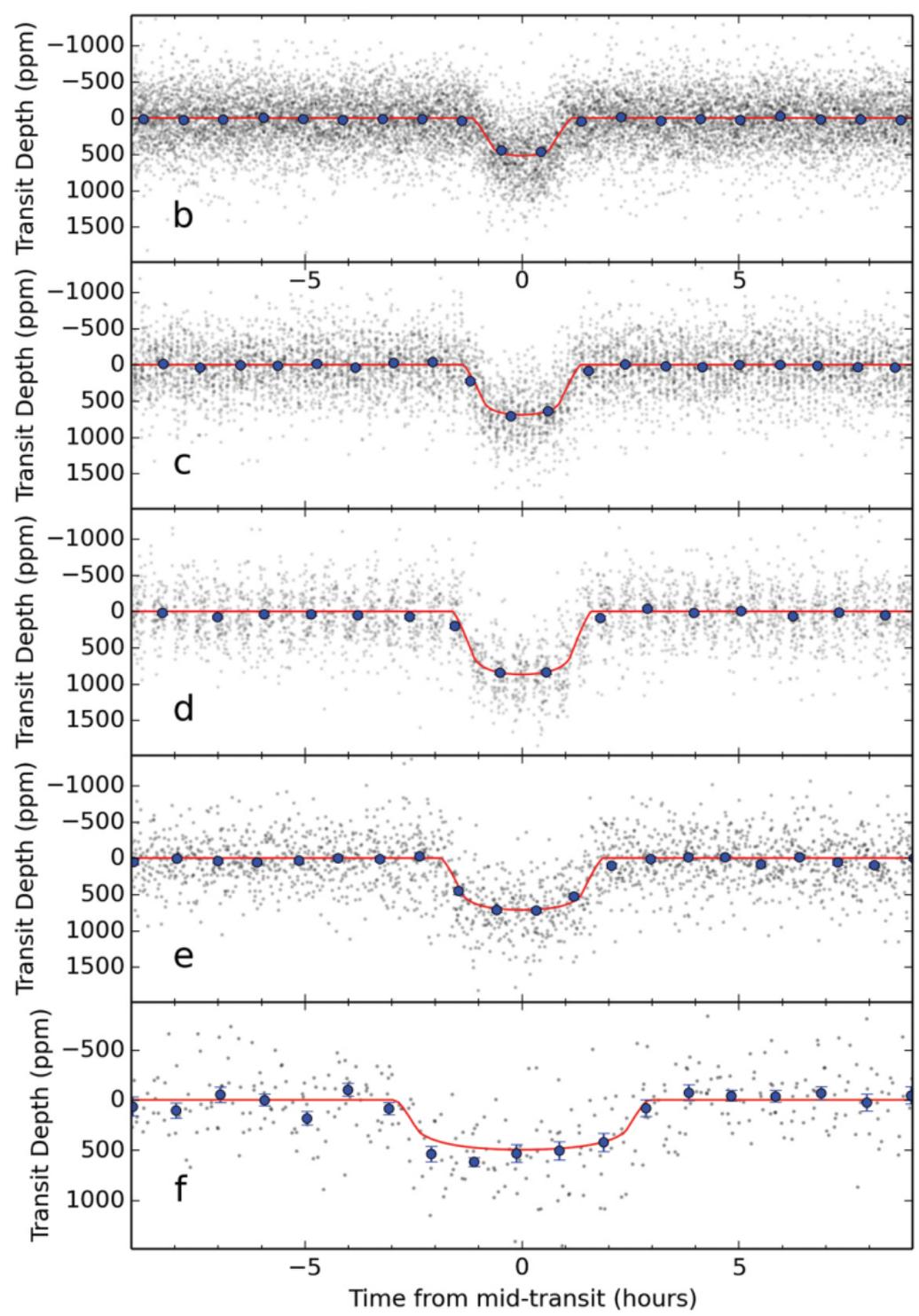

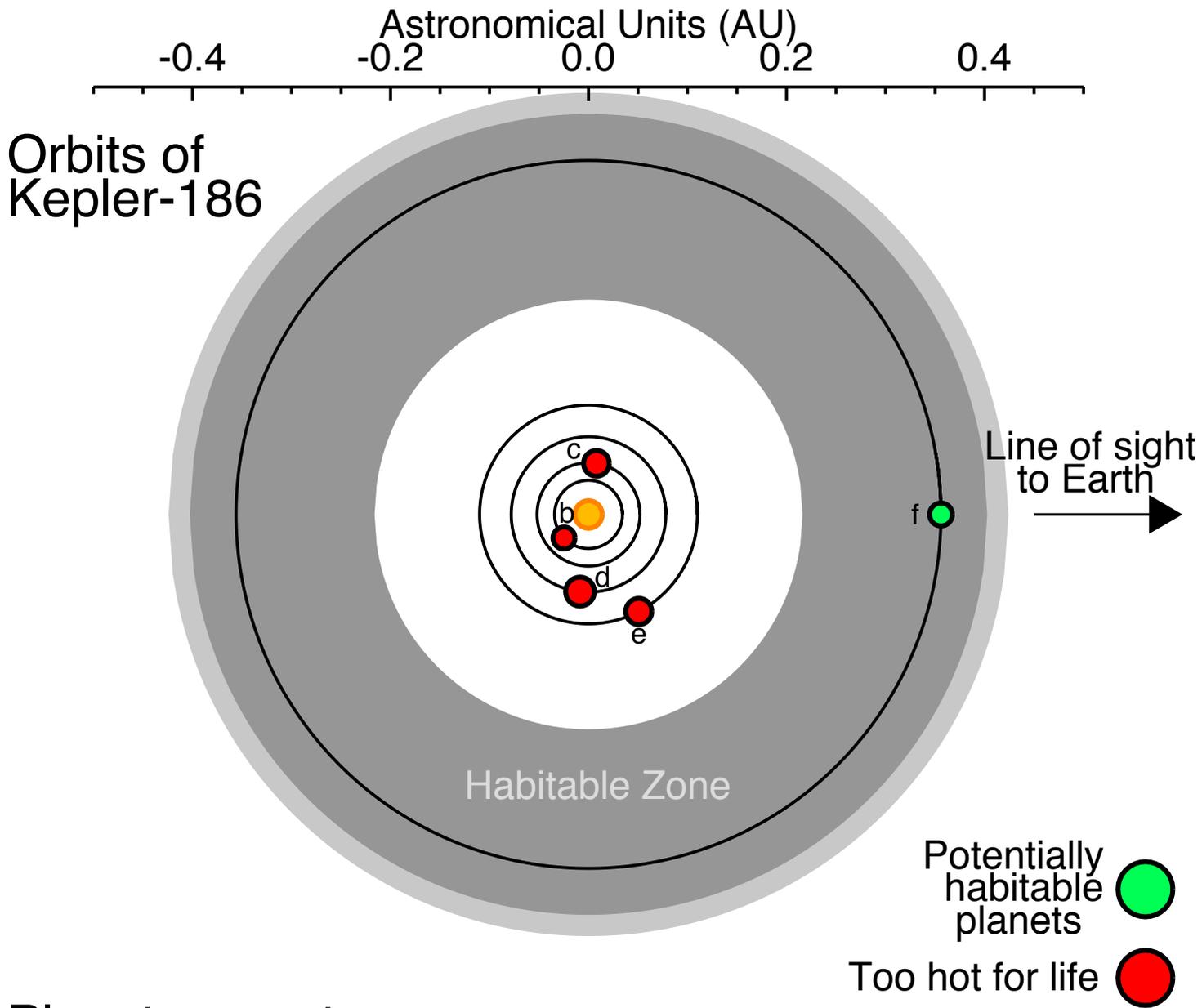

# Supplementary Material

# Contents
1. **Alternative designations, celestial coordinates and apparent magnitudes**
2. **Stellar properties**
3. **Data preparation and transit modeling**
4. **Kepler data validation**
5. **Follow-up observations**
6. **False positive analysis**
7. **Coplanarity**
8. **Orbital stability**
9. **Formation**

## 1. Alternative designations, celestial coordinates and apparent magnitudes
Kepler-186 has the Kepler Input Catalog (KIC) designation 8120608 and coordinates RA=19:54:36.65 and Dec.=45:57:18.1 (J2000). The Kepler project has designated this star Kepler Object of Interest (KOI) 571. Planets b-f have KOI numbers KOI 571.03, .01, .02, .04 and .05. KOI numbers are assigned chronologically with discovery date, hence the inner planet (Kepler-186b) was discovered after planets c and d. The star has a brightness in the Kepler bandpass of $Kp$=14.625, Sloan magnitudes of g=16.049, r=14.679 and i=14.015 and infrared magnitudes of J=12.473, H=11.823 and K=11.605.

## 2. Stellar properties
Kepler-186 was observed as part of a spectroscopic campaign to characterize the cool KOIs (*8*) using the TripleSpec Spectrograph on the 200-inch Hale Telescope at Palomar Observatory. Effective temperature and metallicity were measured using the equivalent widths of the Na I and Ca I lines, as well as by measuring the H20-K2 index (*38*). For Kepler-186 the analysis yielded $T_{eff}$ = 3761±77 K and [Me/H] = -0.21±0.11 dex, which we adopted to begin our analysis. From our transit modeling effort described in Section 3, we determine a mean stellar density of 4.9±1.1 g cm$^{-3}$. To derive interior properties (such as radius, mass and luminosity) of the host star we used a grid of Dartmouth stellar isochrones (*39*) interpolated to a step size of 0.02 dex in metallicity. The observational constraints on the temperature, metallicity, and mean stellar density were fitted to the isochrones to derive the best-fitting model and 1-$\sigma$ uncertainties (Table S1).

Interior models of cool main-sequence stars such as Kepler-186 are well known to show systematic differences to empirically measured stellar properties, with models between 0.3-0.8 $M_\odot$ underestimating radii by up to 10-20% (*14-16*). It has been suspected that these differences are due to enhanced magnetic activity in close binary systems, which can inhibit the efficiency of surface convection (*40*) or cause biases in modeling light curves of heavily spotted stars (*41*). While recent observations have indeed revealed evidence for better agreement in long-period, detached eclipsing binary systems (*42*, *43*), angular diameter measurements of single M-dwarfs from optical long-baseline interferometry still show significant discrepancies (*44*). It is therefore unclear whether evolutionary models are adequate to derive accurate radii of M-dwarfs, and it is important to account for these discrepancies for the derived properties of detected planets (*45*).

Figure S1 shows Dartmouth models in a radius versus effective temperature diagram together with the observational 1-$\sigma$ constraints from the mean stellar density and metallicity. The best-fitting model for Kepler-186 is shown in black, and a sample of stars with interferometric temperatures and radii is shown in red. As expected, empirically measured radii at the measured temperature for Kepler-186 are typically higher than the result derived from model isochrones. To take into account these discrepancies, we have added a 10% uncertainty in quadrature to our stellar radius and mass estimate for Kepler-186, yielding our final estimates of $M_\star = 0.48\pm0.05\ M_\odot$ and $R_\star = 0.47\pm0.05\ R_\odot$. Using the empirical $R$-$T_{eff}$ relation (*46*) would result in a radius of 0.51 $R_\odot$ for Kepler-186, which would translate into a radius of 1.19 $R_\oplus$ for Kepler-186f. Despite being slightly larger, such a radius would still place Kepler-186f well within the regime of plausibly rocky planets (*23*). We hence conclude that systematic differences between models and empirical observations for cool stars do not have a significant influence on the main conclusions of the paper.

Table S1. Stellar Characteristics

| Parameter | Median | ± 1-$\sigma$ |
|---|---|---|
| $M_\star$ ($M_\odot$) | 0.478 | 0.055 |
| $R_\star$ ($R_\odot$) | 0.472 | 0.052 |
| Me/H (dex) | -0.28 | 0.10 |
| $T_{eff}$ (K) | 3788 | 54 |
| $L_\star/L_\odot$ | 0.0412 | 0.0090 |
| log $g$ (cm$^2$ s$^{-1}$) | 4.770 | 0.069 |
| Distance (pc) | 151 | 18 |

Note: The temperature and metallicity were initially derived from K-band spectroscopy and combined with the mean stellar density that we measured from the transit model to derive the interior properties of the star using a Monte Carlo simulation that utilized Dartmouth stellar isochrones. The distance was calculated assuming a line-of-sight extinction of 0.04 magnitudes in the J-band.

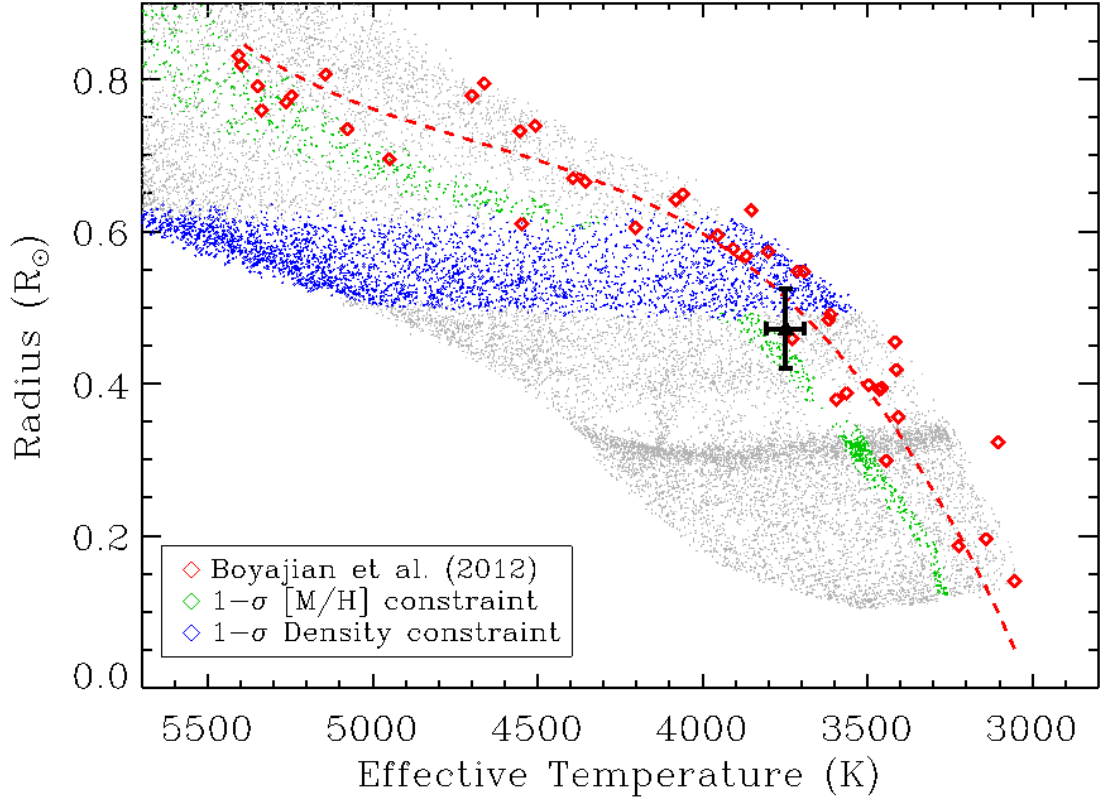

**Figure S1**. Stellar radius versus effective temperature for Dartmouth isochrones with metallicities ranging from -2.5 to +0.5 dex (grey). Green and blue points mark models within 1-$\sigma$ of the spectroscopic metallicity and transit-derived mean stellar density, respectively. The black error bar indicates the derived position for Kepler-186. Red diamonds are interferometric measurements of single stars together with an empirical $R$-$T_{eff}$ relation (dashed line) (*46*).

## 3. Data preparation and transit modeling

We began our analysis of the Kepler observations using simple aperture photometry data (SAP_FLUX) contained in the light curve FITS files hosted at the MAST archive. These data contain both astrophysical variability and uncorrected instrumental systematic noise. We removed most instrumental signals by fitting 'cotrending basis vectors' (*47, 48*) (available from the MAST) to the Kepler time series data using the PyKE software (*49*). Given our goal was to characterize the planets, we removed astrophysical variability (mainly from the rotation of the star) and the remaining instrumental signal using a second order Savitzky-Golay filter with a window of 2 days. The planet transits were weighted zero in this filtering and we treated each observing Quarter of data independently (each Quarter typically includes about 3 months of data sampled at near-continuous 29.4 minute intervals). Finally we normalized the data and combined separate Quarters into a single time series, using Quarters Q0 - Q15 in our analysis.

Our transit model for Kepler-186 consists of five planets with transit profiles calculated using an analytic transit prescription (*11*) with a quadratic limb darkening law. The transit model parameters we sample are mean stellar density ($\rho$), photometric zeropoint, the two limb darkening parameters, a linear ($\gamma_1$) and quadratic term ($\gamma_2$), and for each planet: the midpoint of transit (T0), orbital period ($P_{orb}$), impact parameter (b) and eccentricity vectors $e\sin\omega$ and $e\cos\omega$, where $e$ is eccentricity and $\omega$ is the argument of periastron. We also include an additional systematic uncertainty term ($\sigma_s$) as a model parameter that is added in quadrature with the quoted uncertainty in the Kepler data files ($\sigma_e$).

In the MCMC modeling the photometric zeropoint was assigned an unconstrained uniform prior as were the orbital period of the planets and the time of first transit. The prior on the impact parameter was uniform between zero and (1+k) where k is the planet-to-star radius ratio and the prior on k was uniform between zero and 0.5. Parameterizing $e$ and $\omega$ in terms of $e\sin\omega$ and $e\cos\omega$ enforces a linear prior on $e$ (*50*). While the underlying eccentricity distribution of planets is poorly constrained, it is very unlikely that planets are preferentially in highly eccentric orbits. We assume a uniform prior in $e$ which leads us to include a $1/e$ term as a prior to counteract the bias. The mean stellar density was assigned a Gaussian prior with mean and uncertainty constrained by the spectroscopic observations. For Kepler-186 the model stellar density is not strongly constraining allowing the data to dominate over the prior. Finally, the two limb darkening coefficients are assigned a Gaussian prior with expectation values computed by trilinearly interpolating over ($T_{eff}$, log(g) and Fe/H) tabular data derived with a least-squares method for the Kepler bandpass with Atlas model atmospheres (*51*). The width of the prior on the limb darkening coefficients was taken to be 0.1. We restricted the limb darkening to physical values (*52*). The Gaussian log-likelihood used function was

$$\log \mathcal{L} = -\tfrac{1}{2} \{J\log 2\pi + \log[\sum_{j=1}^{J}(\sigma_{e,j}^2 + \sigma_s^2)] + \sum_{j=1}^{J} \frac{(x_j - \mu_j)^2}{\sigma_{e,j}^2 + \sigma_s^2} \}$$

where $x_j$ is the jth data point in the flux time series with J total observations and $\mu$ is the model. We calculate a log-likelihood to help with numerical stability.

The affine invariant MCMC algorithm we apply involves taking N steps in an ensemble of M walkers and jump n is based on the n-1 position of the ensemble of walkers. We utilized 800 walkers taking 20,000 steps each for a total of 16 million samples.

Parameters derived from the marginalized posterior distributions of the parameters in our MCMC analysis are shown in Table S2. We found a mean stellar density of $4.92^{+0.89}_{-1.06}$ g cm$^{-3}$ and limb darkening coefficients of $\gamma_1 = 0.295^{+0.077}_{-0.077}$ and $\gamma_2 = 0.461^{+0.097}_{-0.094}$.

We then derive from the Markov-chains and the probability distribution of the stellar parameters additional physical characteristics of the planets. The planetary radius is calculated from the posterior distribution of $R_p/R_\star$ multiplied by a normal distribution describing the stellar radius. The semi-major

axis is derived using the formula

$$a = \left(\frac{P^2 \, G * \rho_\star}{3\pi}\right)^{1/3} R_\star$$

where P, ρ and G (the gravitational constant) are in consistent units. The probability distribution of the semi-major axis, a, is computed element-wise in the above equation using the Markov-chain arrays. Finally, the insolation can be calculated independent of the stellar radius. The above equation can be manipulated to be in terms of $a/R_\star$, then insolation (in Earth-Sun units) can simply be derived from

$$S = \left[(a/R_\star)^{-2} \, T_\star^4\right] / \left[(a_\oplus/R_\odot)^{-2} \, T_\odot^4\right].$$

This derivation of insolation keeps all the correlations between parameters from the transit model intact and is not affected by uncertainties in the stellar luminosity and radius.

Table S2. Transit analysis (Median, +/- 1 σ uncertainty)

|  | b | c | d | e | f |
|---|---|---|---|---|---|
| Mid-transit Epoch T0 (BJD-2454833) | 133.3304 +0.0013 -0.0013 | 174.3142 +0.0012 -0.0013 | 176.9045 +0.0014 -0.0015 | 153.8006 +0.0024 -0.0024 | 176.8183 +0.0064 -0.0068 |
| Orbital Period $P$ (days) | 3.8867907 +0.0000062 -0.0000063 | 7.267302 +0.000012 -0.000011 | 13.342996 +0.000025 -0.000024 | 22.407704 +0.000074 -0.000072 | 129.9459 +0.0012 -0.0012 |
| Impact parameter $b$ | 0.30 +0.20 -0.20 | 0.28 +0.20 -0.19 | 0.36 +0.19 -0.24 | 0.31 +0.20 -0.20 | 0.43 +0.19 -0.27 |
| $R_p/R_\star$ | 0.02075 +0.00055 -0.00045 | 0.02424 +0.00056 -0.00047 | 0.02715 +0.00079 -0.00056 | 0.02465 +0.00065 -0.00055 | 0.02144 +0.00103 -0.00092 |
| $e \cos \omega$ | -0.00 +0.23 -0.24 | -0.00 +0.23 -0.24 | -0.00 +0.23 -0.25 | 0.00 +0.24 -0.24 | -0.00 +0.30 -0.34 |
| $e \sin \omega$ | -0.04 +0.07 -0.17 | -0.03 +0.07 -0.14 | -0.03 +0.07 -0.17 | -0.03 +0.07 -0.16 | -0.01 +0.11 -0.21 |
| $R_p$ ($R_\oplus$) | 1.07 +/-0.12 | 1.25 +/- 0.14 | 1.40 +/-0.16 | 1.27 +0.15 -0.14 | 1.11 +0.14 -0.13 |
| Semimajor axis $a$ (AU) | 0.0343 +/- 0.0046 | 0.0520 +/- 0.0070 | 0.0781 +/- 0.010 | 0.110 +/-0.015 | 0.356 +/- 0.048 |
| Insolation $S$ ($S_\oplus$) | 34.4 +6.3 -4.2 | 14.9 +2.7 -1.8 | 6.6 +1.2 -0.8 | 3.33 +0.61 -0.41 | 0.320 +0.059 -0.039 |

Note: The values reported are the median and the central 68% of the probability density. The median values are not intended to be self consistent but represent our knowledge of a parameter's distribution.

## 4. Kepler data validation

The first three planet candidates in this system, Kepler-186b-d, were detected in the first 4 months of Kepler data that include Quarters Q0 - Q2 (*53*). A 4th candidate, Kepler-186e, was detected in the Q1-Q6 data (*54*), and all four of these inner planets were confirmed using Q1 - Q8 data (*9, 10*).

Kepler-186f was detected in the Q1-Q12 data set (*74*), but we used Q1 - Q15 data for our modeling.

Each planet candidate was individually examined to exclude obvious background eclipsing binary induced false positives (*55*). This included looking for differences in the depth of odd and even numbered transits, looking for shifts in the photo-center of the star during the transit and searching for secondary eclipses. All of the candidate planets orbiting Kepler-186 passed the vetting tests.

## 5. Follow-up observations

We undertook an extensive campaign to collect high-contrast images of Kepler-186 in order to establish whether a low mass binary companion to the primary was detectable or if there was a chance alignment of a field star with Kepler-186.

Kepler-186 was observed using the Differential Speckle Survey Instrument (DSSI) on the WIYN 3.5-m telescope on 20110911 in 692-nm and 880-nm filters (approximately *R* and *I*-band). No companions were detected between 0.2-2.0 arcsec. The 5-$\sigma$ detection limit at 0.1 arcsec was 3.6 mag fainter than the target (throughout we will refer to the detection limit at the $\Delta$mag). Kepler-186f was again observed with the DSSI instrument on 20130725, this time using the 8-m Gemini North telescope in the same filters as on the WIYN 3.5-m. Conditions were not optimal on the night when these observations were taken with high cirrus clouds limiting the contrast ratio at 0.2 arcsec to $\Delta$mag=4.9 at 5-$\sigma$. No sources were detected within 0.03-2.0 arcsec of the target star.

On 20130624 Kepler-186 was observed using the natural guide star adaptive optics system with the NIRC2 camera on the Keck-II telescope. A series of *Ks*-band images were obtained using a three-point dither pattern. No nearby sources were detected between 0.2-5.0 arcsec of the target with a $\Delta$mag of 6.9 at 0.5 arcsec.

In Figure S2 we show the regions of parameter space that can be excluded based on speckle observations (left panel) and AO from Keck-II (right panel). Figure S3 shows all the parameter space that can be excluded for each planet candidate with high-contrast imaging constraints converted to the Kepler bandpass (*56*). The regions of $\Delta K$p-separation space where a false positive star cannot exist based on Kepler and follow-up imaging data are shaded green (speckle), pink (AO), blue (transit model) and yellow (Kepler centroid) . The remaining parameter space, shown in white, cannot be excluded and must be accounted for in our false positive calculations.

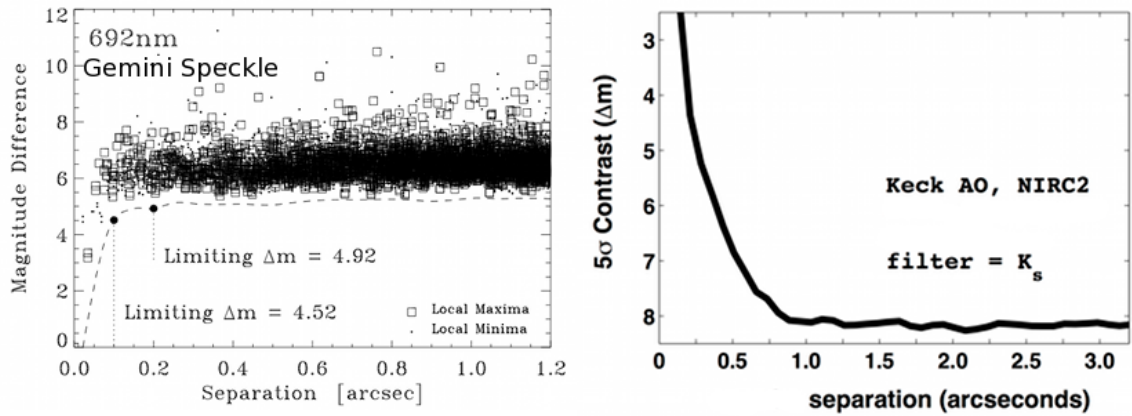

**Figure S2.** Ground-based follow-up observations. Speckle imaging data taken from the WIYN telescope is shown in the left panel and adaptive optics data from Keck II is shown on the right. Each panel shows the limiting magnitude difference as a function of separation where a false positive can be ruled out.

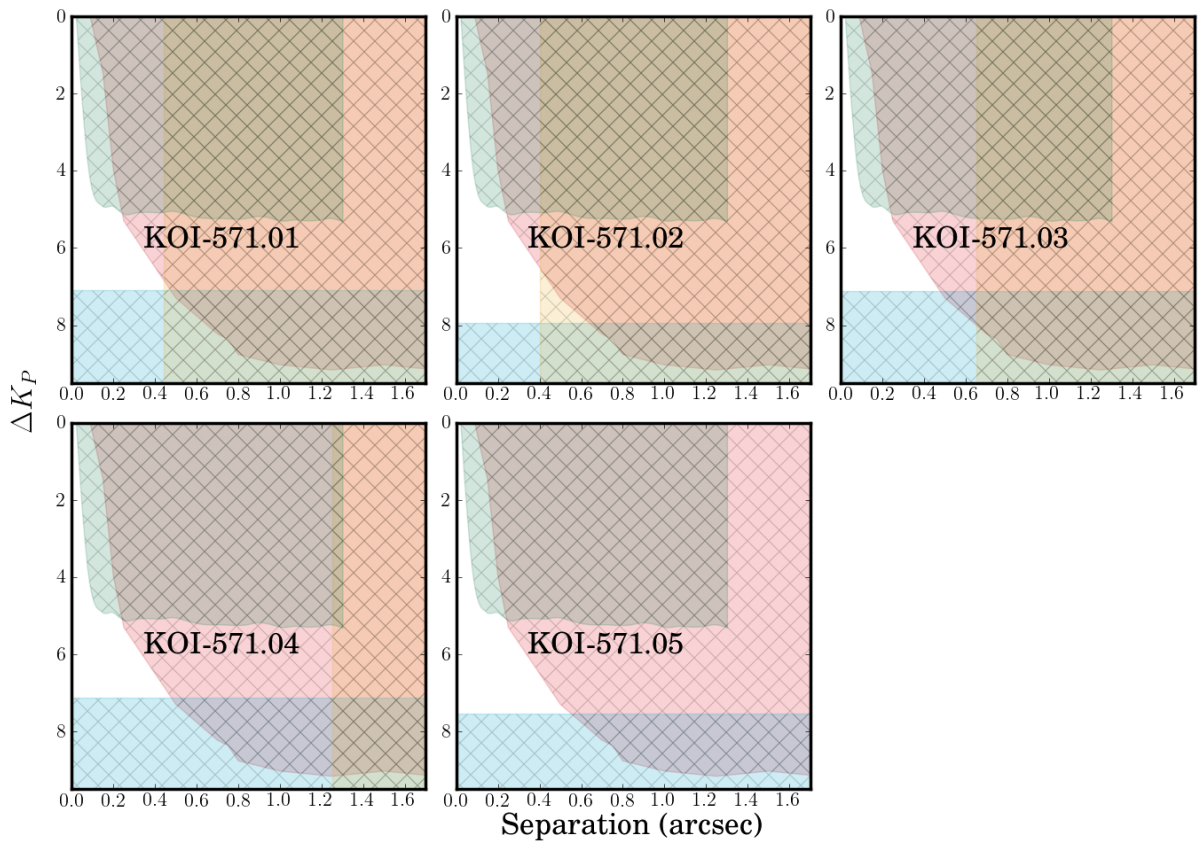

**Figure S3.** Exclusion zones for each of the planet candidates in the Kepler-186 system. Observational constraints rule out false positive inducing stars in all parts of the parameter space save for the white region. High-contrast imaging constraints have been transformed to apply to the Kepler bandpass.

We did not seek radial velocity observations, because a detection of a planet around a star this faint is beyond the capabilities of the current generation of radial velocity instruments.

**6. False positive analysis**

Given the available Kepler and follow-up observations, our goal was to determine the probability that the transit signals we detected were not transits of a planet across the face of Kepler-186. In this analysis we only consider Kepler-186f because the inner four planets have recently been validated as bona fide planets (*9, 10*).

Significant sources of false positives are background or foreground eclipsing binaries, planets orbiting a background or foreground star and planets orbiting a stellar binary companion to the target star. We first consider the case of a background eclipsing binary and background planetary systems. In this statistical analysis we will initially assume that Kepler-186f is the only candidate in the system and then apply a boost to the probability to account for the fact that false positives are significantly less common in multi-planet candidate systems.

We simulate the stellar population in a 1 square degree around Kepler-186 using the TRILEGAL galaxy model (*57*). From this we estimate the stellar density to be 8.8 million stars per square degree brighter than $Kp=32$ in a cone around Kepler-186. As shown in figure S3, we are able to exclude much of the available parameter space that background stars may reside in. We integrated the white region shown in the KOI-571.05 panel of Figure S3 with respect to the Galaxy model to arrive at an estimate of 0.01 stars hidden behind or in front of Kepler-186 that could in principle have a stellar or planetary companion that causes the transit-like signal we observe. We multiply this by the occurrence rate of non-contact eclipsing binaries or planets as seen in Kepler data (2.6%) to arrive at a final probability that the Kepler-186f transit is caused by an eclipsing binary of $2.6\times10^{-4}$. There were 145,000 dwarf stars observed with Kepler for transits but we can only find transit-signals the depth of Kepler-186f in 89% of these stars. Therefore, we estimate the rate of eclipsing binary or background planetary systems in systems such as this will occur 33 times in the Kepler data set. The false positive probability found via computing the ratio of the false positive probability to the a priori probability that Kepler-186f is a planet, found via looking at the number of Earth-sized planets found in Kepler data. To keep our calculations of the multiplicity boost valid, our planet prior probability is the number of Earth-sized planets in single planet systems found in the Q1-12 search of the Kepler data. There were 299 Earth-size planet candidates ($0.8<R_\oplus<1.25$) found in Q1-Q12 data, however, we expect 12.3% of these to be false positives (*58*) leaving 262 Earth-sized planets. Therefore, if Kepler-186f were a single planet system the probability that it were a false positive would be [33/(262+33)] = 11.1%. These odds drop dramatically if we consider the multiplicity of the system. Accounting for multiplicity, the relative probability of a false positive drops by a factor of about 30 (*59*), yielding a confidence in the planetary interpretation of 99.5%.

The second false positive scenario we consider is that of a planet orbiting a stellar companion to the primary star. Typically this scenario is the dominant source of false positives for small planets (*58, 59*). However, with multi-planet candidate hosting M-star primaries such as Kepler-186, we gain a strong constraint on the star that the planets' orbit from transit modeling. Here we assume that all planets orbit the same star but that star's properties are unknown. We repeated the transit modeling described in S3 but removed the mean stellar density and limb darkening priors (save for preventing them from exploring into unphysical parameters (*52*)) and diluted the transit depth appropriately to account for the light of the primary star. This allowed us to place an upper limit on the density of the star hosting the transits of 11.2 g/cc. Interpolating this onto Dartmouth isochrone models, we find a lower limit on the mass of the planet hosting star of 0.39 $M_\odot$. Such a star would only be 0.5 mag fainter than the primary star.

A companion star to Kepler-186 would have to be within a projected distance of 4.2 AU of the primary otherwise our Gemini-North speckle observations would have detected the star. However, if the outer planet was orbiting the secondary it would only be dynamically stable if the closest approach of the stars was not less than 1.4 AU (*22*). We created a population of possible binary companions to the primary using a Monte Carlo simulation method. The distributions of binary separation, mass ratio and eccentricity were taken from the field star statistics (*60*). We adopted a binarity rate of 10% which is appropriate for exoplanet hosting binaries separated by less than 10 AU (*21*). From the sample of potential companions we removed all companions that would not be allowed based on observations or would cause the outermost planet to become unstable from dynamical interactions from the brighter star. We ran our simulations $10^7$ times and an undetected companion was found on 445 occasions, an occurrence rate of $4.4 \times 10^{-5}$. The planet may still orbit the primary star, however, in the 445 cases with a feasible companion. We assigned a likelihood that the planet orbits the companion star that is proportional to the ratio of the probability to transit for the secondary relative to the primary star (a planet is less likely to transit a smaller star). This yields an occurrence rate of this false positive scenario of $2.1 \times 10^{-5}$ per star observed by Kepler. An estimate of the occurrence rate in Kepler data of the scenario whereby the planet orbits a companion star can be estimated by multiplying the false positive probability per star by the number of stars hosting planet candidates, i.e. ($2.1 \times 10^{-5} \times 3601$) = 0.075 false positives. Compared to our a priori estimate of 417 for the occurrence rate of stars hosting at least one Earth-sized planet (there are 475 stars with at least one Earth-sized planet and we assume a false positive rate of 12.3%) our confidence that Kepler-186f orbits the primary star rather than an unseen companion is 99.98%.

### 7. Coplanarity

The relative durations of planets in a multi-planet candidate system can also provide information on whether all the planets orbit the same star. For a given planet with a central transit (impact parameter *b*

= 0) and circular orbit, the transit duration ($D$) is proportional to $P^{1/3}$ (where $P$ is the orbital period in days). If the scaled durations ($D/P^{1/3}$) of each planet in a system are equal, then it is highly likely that these planets orbit the same star (*61*). The scaled durations for the five planets of Kepler-186, all of which are consistent with having circular orbits, are shown in Figure S4 in order of ascending orbital period. The scaled durations are very similar among the planets, providing support that the signatures in the Kepler-186 light curve are due to five planets orbiting the same star. Interestingly, the scaled durations decrease slightly with planet orbital period, suggesting that they transit at increasing impact parameters. Increasing impact parameters as a function of orbital distance is expected if the planets lie in the same orbital plane, indicating the Kepler-186 planets are relatively coplanar.

Note that in order for the inner planet (b) to transit, we must view the system within a known angle of about 3.7 degrees, and in order for the outer planet (f) to transit, the viewing angle has the smaller range of only about 0.35 degrees. As a result, the planetary orbits are constrained to be nearly coplanar (within a few degrees). In comparison, circumstellar disks typically have scale heights (in the inner regions) of order $h/r \sim 0.05$, which implies co-planarity to about 3 degrees.

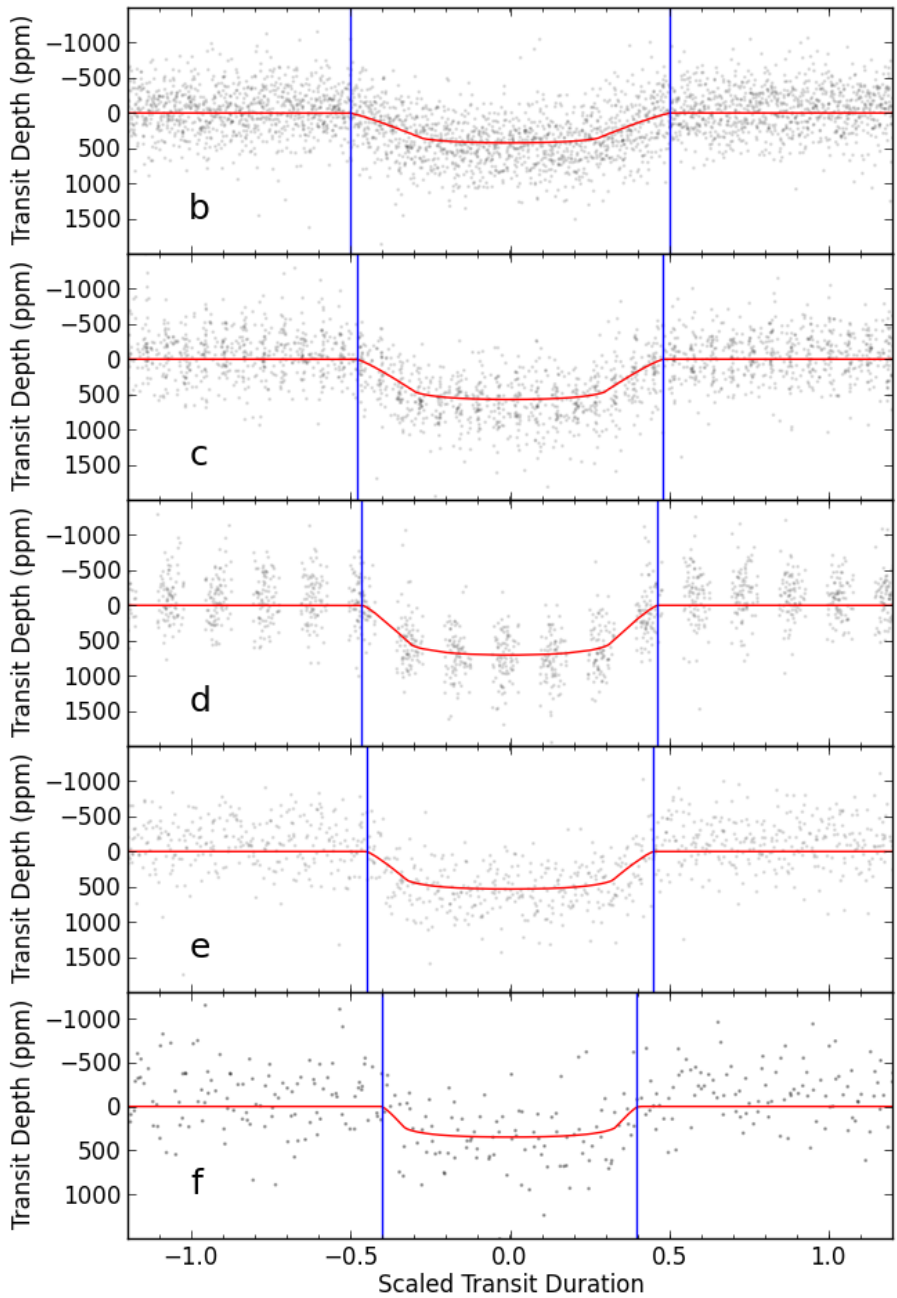

**Figure S4**. Normalized durations of the Kepler-186 planets. Each panel shows the folded light curve of a planet normalized by the cube root of the orbital period ($P^{1/3}$). This allows a direct comparison of the relative durations. If all of the durations are consistent, as they are in this system, it is highly likely that all planets orbit the same star.

## 8. Orbital Stability

A system of two planets in orbit around a star is dynamically stable if the objects are spaced by roughly 3.5 or more mutual Hill radii $R_{H,m}$ (62, 63). The mutual Hill radius is defined as $R_{H,m} = 1/2\ (a_1 + a_2)[(m_1 + m_2)/3M_*]^{1/3}$, where subscripts 1 and 2 refer to the two planets, $a$ is the orbital semimajor axis, $m$ is the planet mass, and $M_*$ is the stellar mass. A system of many planets must be more widely separated than a critical limit of 5-10 mutual Hill radii to ensure long-term stability (64, 65).

Although we don't have mass estimates for the Kepler-186 planets which are needed for our dynamical stability simulations, we assume they are terrestrial in nature using the following reasoning. All five planets around Kepler-186 have sizes less than 1.4 $R_\oplus$, which thermal evolution models predict are too small to be dominated with a low density H/He gas envelope (23). Furthermore, any H/He envelope that may have been accreted was likely lost via photoevaporation early in the star's lifetime. With a measured rotation period of 34 days (66, 67) and no significant flaring observed in the Kepler light curve, Kepler-186 is likely older than about 4 Gyrs (68, 69) which is old enough such that the main phase of atmospheric erosion should have already occurred. If we assume a simple model of energy limited escape (3), and scale the UV flux to that of the UV-quiet Sun at the distance of Kepler-186f, then the estimated mass loss rate is about $10^{-4}\ M_\oplus$/Gyr. As a result, in a few Gyr, only a relatively small fraction of the planet's mass would be evaporated, but this change in mass corresponds to roughly 100 times the current mass of the atmosphere of Earth. These planets are much more likely to be composed of some combination of denser material (silicate rock, iron, and/or water). Table S3 lists the planets' masses for a wide range of plausible compositions using widely-used theoretical mass-radius relations (70).

Table S3. Range of plausible planet masses (in $M_\oplus$)

|  | Pure Ice ($H_2O$) | Pure Rock (Silicate) | Earth-like | Pure Iron |
|---|---|---|---|---|
| Planet b | 0.28 | 0.90 | 1.26 | 3.23 |
| Planet c | 0.46 | 1.56 | 2.24 | 6.30 |
| Planet d | 0.66 | 2.38 | 3.50 | 10.00 |
| Planet e | 0.48 | 1.66 | 2.38 | 6.76 |
| Planet f | 0.32 | 1.02 | 1.44 | 3.77 |

The dynamical inter-planet spacing depends on the planets' true masses. Planets b and c are dynamically closest together and planets e and f are the most widely spaced. Lower-density, lower-mass planets are more widely spaced in dynamical terms (mutual Hill radii). For pure ice, planets b and c are separated by 12 $R_{H,m}$, but this value decreases with increasing planet mass to 7 $R_{H,m}$ for Earth-like compositions and just 5 $R_{H,m}$ for pure iron planets. The gap between planets e and f is wide enough to fit another planet. For ice-rock-Earth-iron planets, the gap is 45-30-26-19 mutual Hill radii wide.

To examine the dynamical stability of the Kepler-186 system, we ran a suite of N-body simulations of the five-planet system for the full range of planetary compositions. Given the weak constraints on the planets' eccentricities and longitudes of pericenter, we sampled a range of orbital phases and included initial eccentricities up to 0.05. In all cases the systems were stable for the 0.1 Myr duration. We ran 10 longer-term simulations (without tides or general relativity) with pure iron planets, all of which were stable for 100 Myr, the duration of the simulations.

We stress that close-packed multi-planet systems display chaotic dynamics, so that the question of stability must be addressed statistically. As a result, a great deal of additional dynamical work can be done to explore the long term stability of this system. Such work would result in stronger constraints (upper limits) on the (as yet unknown) planetary masses.

## 9. Formation

The planets in our solar system are thought to have formed in situ by accreting local material from a protoplanetary disk of gas and dust that surrounded the newly formed Sun (*71*). The discovery of hundreds of short-period exoplanets and the diversity of their sizes, masses and system architecture has shed new light on classical planet formation theories. Many systems require some form of inward migration of mass during their formation (*28*, *29*), or planet-planet scattering after they formed to bring them inwards, to explain their observed properties.

We performed a suite of N-body numerical simulations of late-stage accretion around a star like Kepler-186 in order to shed light on the types of protoplanetary disks that would be needed to form planetary systems like Kepler-186b-f in situ. We examine populations of 20 planetary embryos and/or 200-400 planetesimals spread between 0.02-0.5 AU. The surface density profile was varied between $r^{-1}$ and $r^{-2.8}$, and the total mass in solids ranged from 6-10 $M_\oplus$. Our simulations begin at the epoch of planet formation just after the gas in the disk is dissipated, and each system was integrated with the *Mercury* hybrid integrator (*72*), which treats collisions as perfect mergers, for 10 Myr using a 0.2 day timestep.

Figure S5 shows the outcome of eight of the *N*-body simulations. Simulations that began with more massive initial disks (10 $M_\oplus$) and steeper surface density profiles ($r^{-2.5}$) were most successful in reproducing the broad mass-orbital radius distribution of the Kepler-186 system. Disks with this much mass so close to their star or with such steep surface density profiles, however, are not commonly observed (*30*), suggesting that the Kepler-186 planets likely formed from material that underwent some inward migration in the earlier stages of formation while gas was still present in the disk, or perhaps were perturbed inwards by a distant planet or star.

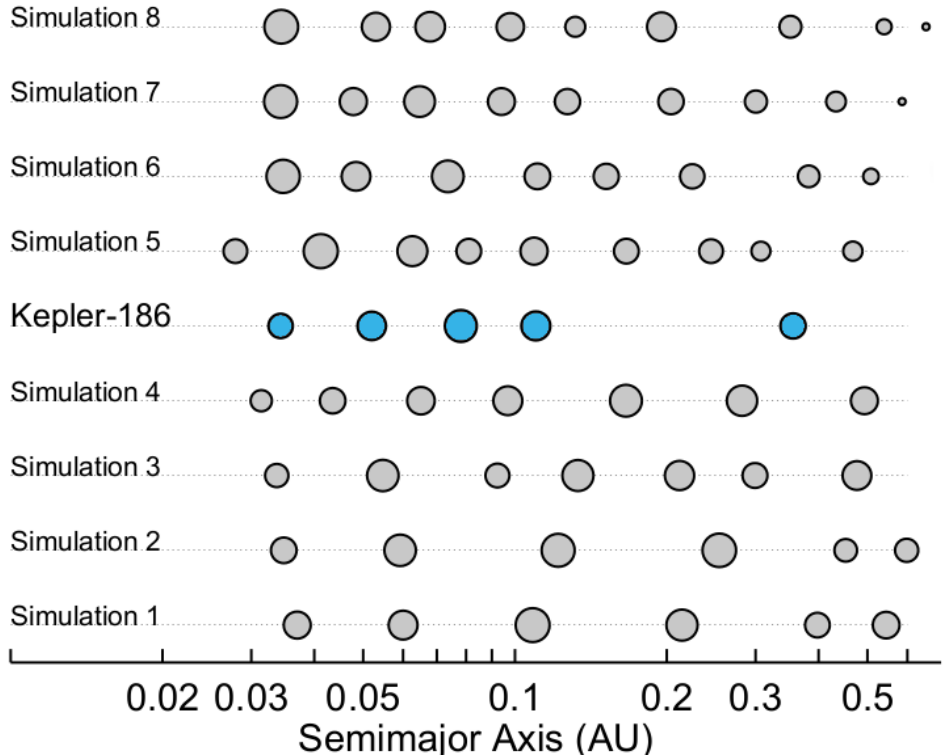

**Figure S5**. Final planetary systems that formed in eight of our simulations. Each symbol represents a final planet and the symbol size is proportional to its radius derived from theoretical mass-radius relations (*73*). The Kepler-186 planets are shown in blue. Simulations 1-4 began with a surface density profile of $r^{-1.8}$ and simulations 5-8 began with a steeper profile of $r^{-2.7}$. All simulations formed 1-2 planets in orbits between those of Kepler-186e and Kepler-186f.

Regardless of how these planets formed, these simulations demonstrate that a sixth planet could in theory remain stable in-between the orbits of known planets e and f (0.15-0.35 AU). Could such a planet exist in the system but not transit? For that to be the case, that planet would need to have a modest inclination of at least a few degrees with respect to the common plane of the other planets. A collision or scattering event after the dissipation of the gaseous disk could produce such an inclination. It would then be a simple coincidence that planet f's orbit is aligned with the inner ones whereas this extra planet's is not. Or, if this extra planet is relatively lower-mass than the other planets then its secular oscillations in inclination can simply reach a higher amplitude than the other planets, decreasing the probability of observing it in common transit with the other planets.